\documentclass[12pt,a4paper]{article}
\usepackage{epsf}
\usepackage{graphicx}
\usepackage{natbib}
\usepackage{amsmath,amssymb}
\usepackage{url}
\usepackage{cite}
\usepackage{xcolor}

\usepackage{journals_joge}

\begin{document}

\title{Preliminary analysis of the annual component of the polar motion over 180-year data interval}

\author{Natalia Miller, Zinovy Malkin\\ Pulkovo Observatory, St. Petersburg 196140, Russia\\ e-mail:natmiller@list.ru}

\date{\vskip -2em}

\maketitle

\begin{abstract}
The paper presents preliminary results of studying variations
in the annual component in the Earth's polar motion.
For this purpose, a signal with an annual period was extracted,
firstly, from the series of pole coordinates of the International
Earth Rotation and Reference Systems Service (IERS), and secondly,
from the combined series of Pulkovo latitude variations for 1840--2017.
For this purpose, one-dimensional and multidimensional singular
spectrum analysis was used.
The Hilbert transform was used to calculate the change in the amplitude
and phase of the annual oscillation over time.
As a result, it turned out that over an interval of about 180 years,
an almost monotonic increase in the amplitude of the annual oscillation
from $\approx$60~mas to $\approx$90~mas and an almost monotonic phase shift
of $\approx$45$^\circ$ are observed.
A correlation was also found between the amplitude of the annual
component and the difference in average temperatures from November
to March in the northern and southern hemispheres.
\end{abstract}


\section{Introduction}

There are two main components in the Earth's polar motion:
the Chandler wobble (CW) with a period of about 14 months and the annual wobble (AW).
In the literature, in most cases, the results of studying the first of them are given,
see, for example,
\citet{Vondrak1988,Nastula1993,Schuh2001PM,Miller2011,ChaoChung2012,Zotov2022MUPB}
and paper referenced therein.

The annual component of the polar motion has been studied much less frequently, especially
based on long data series longer than a hundred years.
As an example, in the works \citet{Vondrak1988,Nastula1993,Schuh2001PM},
variations in the AW amplitude are identified.
All studies show that these changes are much smaller in magnitude
than the variations in the CW amplitude.
At the same time, the results obtained by different authors differ somewhat.
In addition, \citet{Vondrak1988} identified variations in the AW phase,
and in \citet{Schuh2001PM} (mathematically equivalent)
possible changes in the AW period are considered.

Thus, it can be concluded that the AW variations have not yet been sufficiently
investigated and it makes sense to continue these studies
using longer data series and alternative mathematical methods.
The present work is a step in this direction.
Here, we studied significantly longer series of pole coordinates than were used
in previous works, and a unique combined series of changes in Pulkovo latitude
of about 180 years.
In contrast to the above-mentioned works, singular spectral analysis
(SSA) \citep{Golyandina2001} was used to identify the AW component in polar motion.


\section{AW analysis}
\label{sect:}

In this paper, the annual component of the Earth's pole motion was investigated
using the pole motion data of the International Earth Rotation and Reference
Systems Service (IERS):
IERS C01 series for 1846--2018 and IERS C04 for 1962--2018,
as well as combined series of Pulkovo latitude variations for 1840--2014 ($\varphi$),
which directly reflects all changes in the pole coordinates,
see (\ref{eq:delta_phi_xy}).
To form the combined $\varphi$ series,
Pulkovo latitude determinations made in different periods of time on different
instruments of the Pulkovo Observatory were used.
In periods for which Pulkovo latitude determinations are unavailable, the series
is supplemented with values calculated from the pole coordinates $X_p$ and $Y_p$
of the IERS C01 and C04 series using the formula:
\begin{equation}
\Delta\varphi = \varphi-\varphi_0 = X_p \cos\lambda + Y_p \sin\lambda =
0.8631\,X_p - 0.5049\,Y_p \,,
\label{eq:delta_phi_xy}
\end{equation}
where $\lambda$ is Pulkovo latitude.
The data used for the combined Pulkovo latitude series for
different time periods are given in Table~\ref{tab:series}.
The process of constructing a combined latitude series is described in more detail in \citep{Miller2011}.

\begin{table}[ht]
\centering
\caption{Data used to construct the Pulkovo combined latitude series.}
\label{tab:series}
\begin{tabular}{lcc}
\hline
Telescope/series & Data type & Dates \\
\hline
Ertel large vertical cicle                       & latitude     & 1840--1842 \\
Repsold transit instrument in the prime vertical & latitude     & 1842--1846 \\
IERS C01                                         & $X_p$, $Y_p$ & 1846--1904 \\
ZTF-135                                          & latitude     & 1904--1941 \\
IERS C01                                         & $X_p$, $Y_p$ & 1941--1948 \\
ZTF-135                                          & latitude     & 1948--2006 \\
IERS C04                                         & $X_p$, $Y_p$ & 2006--2018 \\
\hline
\end{tabular}
\end{table}

\begin{figure}
\centering
\includegraphics[width=\textwidth]{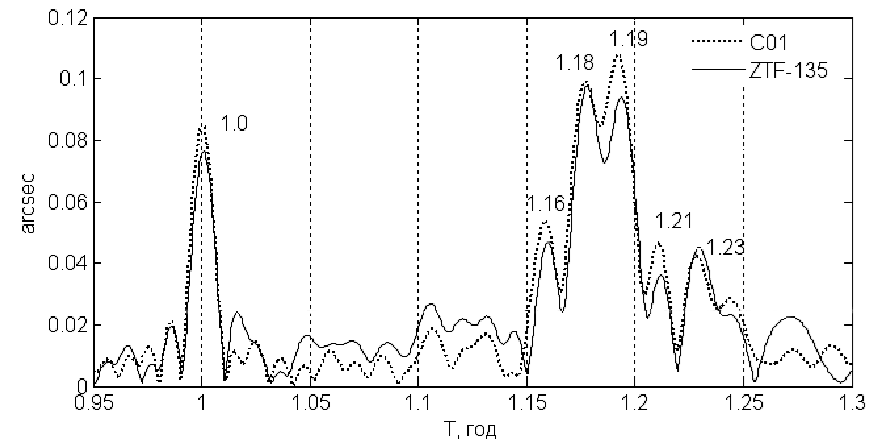}
\caption{Fourier spectra of the IERS C01 and combined Pulkovo latitude series.}
\label{fig:spectrum1}
\end{figure}

Fig.~\ref{fig:spectrum1} shows the spectra of the two main series used in this work:
the IERS C01 series and the combined Pulkovo latitude series.
The spectrum covers a range of periods that includes both the AW and CW.
It can be seen that that these two components can be effectively separated by using
suitable bandpass filtering.

To extract and analyze the AW signal, the SSA method was used in this work.
This method and its multivariate modification (MSSA) are based on the transformation of
a time series into a matrix and its singular value decomposition, which results
in the decomposition of the original series into additive components.
When using this method, a sample correlation matrix is calculated,
whose eigenvalues $\lambda_i$ are the sample variances of the corresponding principal
components.
These components are determined in such a way that the first of them gives the maximum
possible contribution to the total variance.
The performed transformation does not change the sum of the variances, but only
redistributes it so that the greatest variance falls on the first components, which
makes it possible to exclude from the analysis components that have small variances and,
accordingly, a relatively small contribution (relative signal power) to the process
under study.
The percentage contribution of the $i$-th component is calculated using the formula:
\begin{equation}
V_i = \frac{\lambda_i}{M} \times 100\% \,,
\label{eq:labmbda_i}
\end{equation}
where $M=N/2$, $N$~-- the length (number of points) of the series,
$\lambda_i$~-- the $i$-th eigenvalue.

The complex Hilbert transform was applied to determine the variations in the amplitude
and phase of the annual component of the pole motion (calculations were performed
with the \texttt{hilbert} function from the Matlab Signal Processing Toolbox).


\section{AW analysis}
\label{sect:annual_term}

\begin{figure}
\centering
\includegraphics[width=12cm]{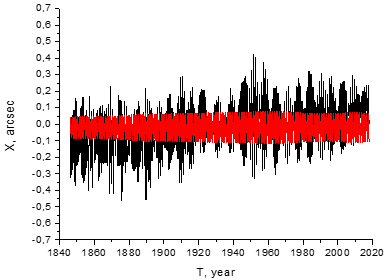}\\[5ex]
\includegraphics[width=12cm]{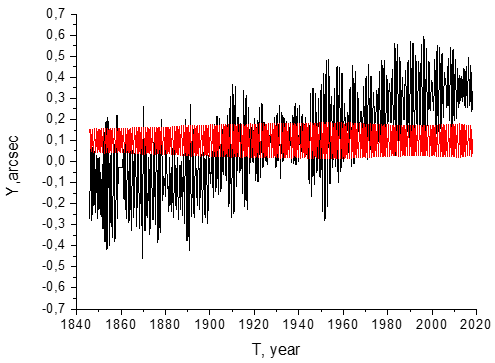}
\caption{Result of applying the MSSA to the IERS C01 series:
top~-- pole coordinate $X_p$, bottom~-- pole coordinate $Y_p$.
Black line shows the total variations of pole coordinates,
red line shows the AW component.}
\label{fig:ssa_xy}
\end{figure}

The AW signal was extracted from the IERS C01 series (black lines in
Fig.~\ref{fig:ssa_xy}) using the MSSA method, which allows
to jointly analyze the series of polar coordinates $Xp$ and $Yp$ as a single
two-dimensional data series.
The resulting annual signal is shown by the red lines in Fig.~\ref{fig:ssa_xy}
together with the original series of IERS C01 polar coordinates (black lines).

\begin{figure}
\centering
\includegraphics[width=\textwidth]{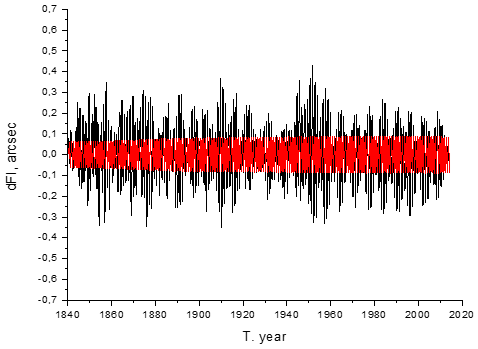}
\caption{Result of applying SSA to the combined Pulkovo latitude series.
Black line shows the total latitude variation, red line shows the AW component.}
\label{fig:ssa_phi}
\end{figure}

An alternative series of the annual signal in the polar motion was extracted from
the 180-year combined Pulkovo latitude series using the one-dimensional SSA version.
This series is shown in Fig.~\ref{fig:ssa_phi}.
The black line in the figure shows the original series of Pulkovo latitude variations.
As can be seen from the comparison of the presented data, all three AW series
are close to each other.

\begin{figure}
\centering
\includegraphics[width=\textwidth]{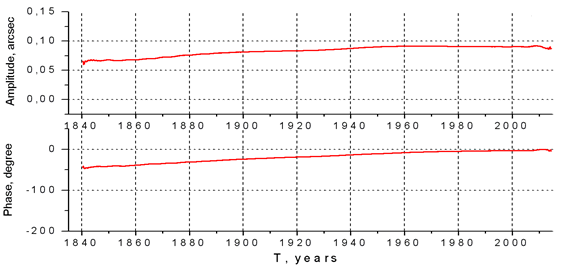}
\caption{The amplitude and phase AW variations extracted by the SSA method from
the combined Pulkovo latitude series. The amplitude and phase
have increased by 0.03$''$ and 45$^\circ$, respectively, over 180 years.}
\label{fig:amplitude_phase}
\end{figure}

A further analysis of the AW series was carried out using the Hilbert transform,
which allows us to study the variations in the amplitude and phase of this oscillation.
The results of this analysis, presented in Fig.~\ref{fig:amplitude_phase},
show that over the 180-year period under study,
there is a slow increase in the amplitude of the annual term from~$\approx$60~mas
to~$\approx$90~mas until the early 1960s,
after which the amplitude remains virtually constant.
A similar behavior is demonstrated by the AW phase, which increased
by~$\approx$45$^\circ$ from 1840 to the early 1960s, after which it began to change
much more slowly.

\begin{figure}
\centering
\includegraphics[width=14cm]{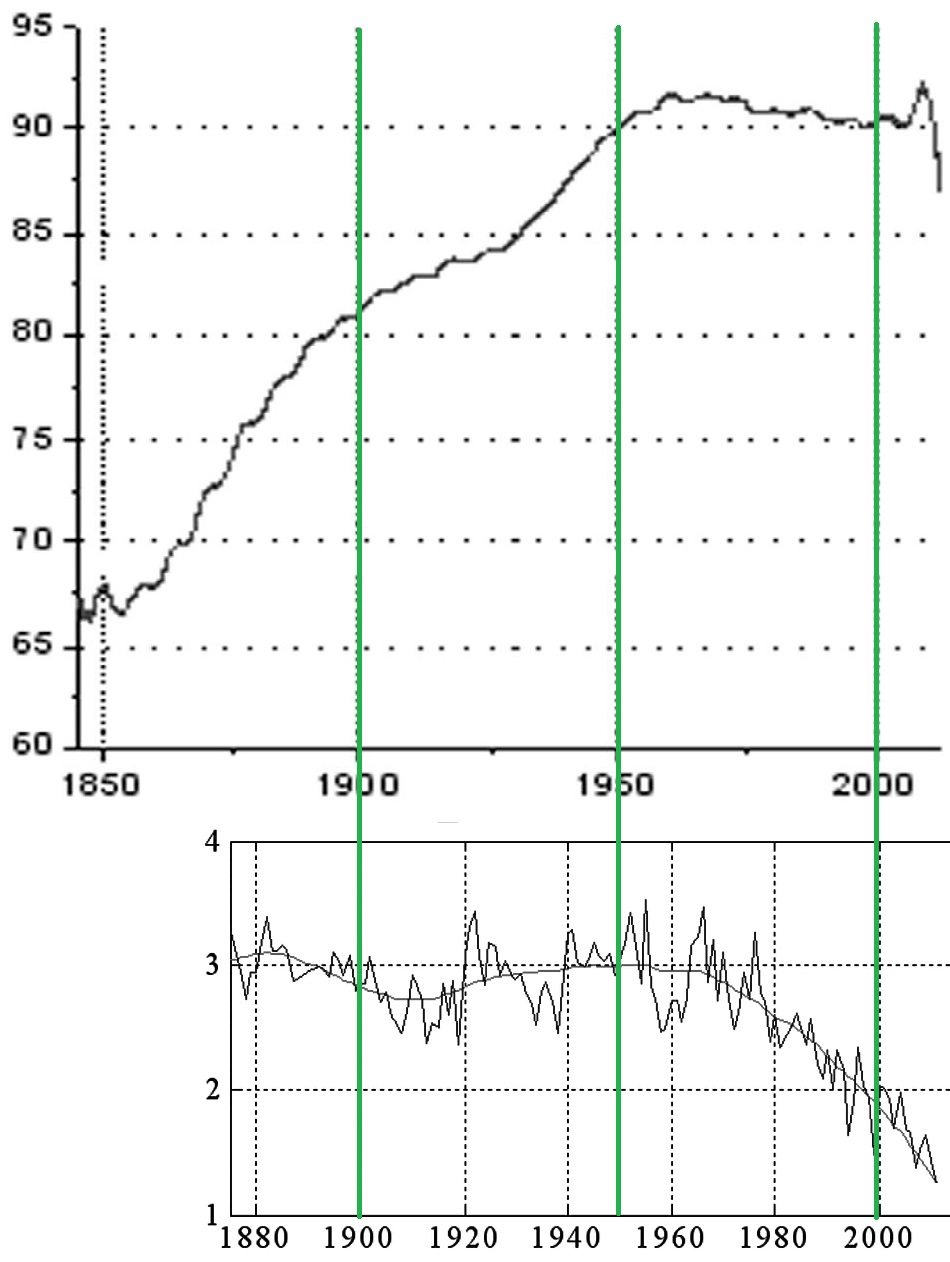}
\caption{Top panel~-- the amplitude of the annual component of the change
in Pulkovo latitude, mas, bottom panel~-- the difference in average temperatures
November--March of the northern and southern hemispheres of the Earth, $^{\circ}$C.}
\label{fig:dfiamp_tempdiff}
\end{figure}

Fig.~\ref{fig:dfiamp_tempdiff} shows a comparison of the latitude change data
with one of the climate data series:
the curve of the annual component amplitude change is shown at the top,
the curve of the difference in average temperatures for November-March in the northern and southern hemispheres
of the Earth is shown at the bottom\footnote{\url{ftp://ftp.cdc.noaa.gov/Datasets/20thC_ReanV2/Monthlies/gaussian/monolevel/}}.
Both curves show an inflection around 1960~. It is known that the annual oscillation
in the polar motion is explained as a forced oscillation caused by seasonal changes
in the atmosphere, ocean, and hydrosphere.
Thus, it can be assumed that one of the causes of this phenomenon
is the impact of climate change on the Earth's rotation through long-term
climatic changes in global atmospheric processes.

\clearpage
\section{Conclusions}
\label{sect:conclusions}

In this paper, a preliminary study of the annual component of the polar motion was carried
out using IERS series C01 and the combined Pulkovo latitude series over a period of 180 years from 1840 to 2018.
Using the SSA method, the annual component and variations in its amplitude and
phase were extracted and analysed from these series.
A comparison of the parameters of the variation in the annual component of the pole motion
calculated from the two original data series showed that they were very close to each
other.
As a result, it was found that a number of parameters of the annual component of the pole
motion over an interval of about 180 years demonstrate an almost monotonic increase in
amplitude from $\approx$60~mas to $\approx$90~mas with a simultaneous monotonic phase
shift of $\approx$45$^\circ$.
At the same time, the increase in amplitude and the phase shift practically ceased
about 60 years ago.
Also, the variations in the annual component show features
in the behavior of its amplitude near the period of the minimum amplitude of the Chandler
oscillation in the 1920s.
A correlation was also found between the amplitude of the annual component
of the polar motion and the difference in average temperatures from November to March
in the northern and southern hemispheres.
This allows us to assume a connection between the parameters of the Earth's polar motion
and climate change, which may be a reflection of the
influence of various processes in the atmosphere and hydrosphere on the polar motion.


\section*{Acknowledgments}

\bibliography{references_eng.bib}
\bibliographystyle{joge}

\end{document}